      [1999/12/01 v1.4c Il Nuovo Cimento]
\documentclass[varenna]{cimento}

\newcommand{\beq}[1]{\begin{equation}\label{#1}}
\newcommand{\eeq}{\end{equation}}
\newcommand{\eqref}[1]{(\ref{#1})}

\newcommand{\qb}{{\bf q}}
\newcommand{\gb}{{\bf g}}
\newcommand{\ub}{{\bf u}}
\newcommand{\vb}{{\bf v}}
\newcommand{\Fb}{{\bf F}}
\newcommand{\rb}{{\bf r}}
\newcommand{\Rb}{{\bf R}}
\newcommand{\sigb}{{\bf \sigma}}
\newcommand{\pQ}{{\cal Q}}

\usepackage{graphicx}  

\title{Elastic properties of  colloidal solids with disorder}

\author{Matthias Fuchs}

\institute{Fachbereich Physik, Universit\"at Konstanz, 78457 Konstanz, Germany} 

\begin{document}

\maketitle

\begin{abstract}
Recent progress in  approaches to determine the elastic constants of solids starting from the microscopic particle interactions is reviewed. 
On the theoretical side, density functional theory approaches are discussed and compared to more classical ones using the actual pair potentials. On the experimental side, video microscopy has  been introduced  to measure the elastic constants  in colloidal solids. For glasses and disordered systems, the theoretical basis is given for this novel technique, and some challenges and recent advances are reviewed.
\end{abstract}

\section{Introduction}

Solids respond elastically to deformations. Hooke formulated the laws of linear elasticity as early as 1678 in the anagram  'ceiiinosssttuv' which stood for 'ut tensio sic vis'. Solid state physics developed the atomistic view of ordered solids, starting in the classical realm and  leading to quantized lattice vibrations \cite{born}. Non-equilibrium thermodynamics focused on  the fundamental principles at work in elasticity \cite{Martin1972}.    Spontaneous breaking of a continuous symmetry leads to equilibrium states, which are characterized by finite order parameters and  are degenerate in energy. Homogeneous fluctuations, which restore the symmetry and transform one ordered state into another equivalent one, require no energy. If these fluctuations are spatially inhomogeneous, they cause a small free energy increase. It is captured in the elastic free energy which depends on the gradients of the fluctuations. Elastic constants or elastic moduli appear as second order derivatives of the free energy. They can be measured in the long wavelength limit of the dispersion relations, which contain the eigenvalues of the dynamical matrix as function of the wavevector. The elastic moduli are also connected to the correlation functions of the symmetry restoring fluctuations (the so-called Nambu-Goldstone modes), which are long ranged.    In ordered solids, translational and rotational invariance of the underlying Hamiltonian function of the particles is broken, and (transverse acoustic) sound waves are the Goldstone modes. This general framework is well understood and discussed in textbooks like Ref.~\cite{Chaikin95}, and many important insights are possible even with simple and intuitive considerations using  mean field theory.  

Colloidal dispersions have recently  moved into the foreground for studying elastic phenomena. On the one hand, the rheology of colloidal dispersions is a well established area, and has provided important information on the macroscopic viscoelastic behavior.  Rheology, including the recently introduced micro-rheology where a probe particle is strongly forced through a sample, is especially aimed at studying the non-linear response. Plastic, an-elastic, and yielding behavior can be investigated \cite{Siebenbuerger2012}. On the other hand, microscopy of particle trajectories has recently become possible and provides detailed information on the local motion. The size and time scales of colloidal particles are the basis for this unique possibility. Thermal fluctuations can be recorded and spatially resolved dynamics   investigated \cite{Keim2004,Reinke2007}. Rare events like the ones  dominating nucleation can be observed. This review is concerned with the theoretical basis for using video and confocal microscopy for determining elastic properties in colloidal solids, especially in crystals with defects, and in amorphous solids.  

The elastic energy contributions, the corresponding moduli, and  the elastic field become complicated concepts if disorder is present. Topological defects destroy the order and are an important topic with well established consequences on material properties; they are not discussed in this mini-review, but are covered extensively in Ref.~\cite{Chaikin95}. Point defects like interstitials and vacancies do not destroy the long-ranged order but lead to an additional hydrodynamic field, the defect density. While its appearance and coupling to the other elastic fields had been well understood from non-equilibrium thermodynamics point of view  \cite{Martin1972}, the  microscopic formulae  for all elastic constants in the presence of defect densities had not been known. Recently, they were established within density functional theory \cite{Walz2010}, as will be reviewed here.  The effect of defect diffusion on dispersion relations will be discussed, if defects are falsely ignored or taken into account properly.  As an introduction, a short summary of the more classical approach based on the particle pair potential will be given. The second part of this short review then concerns the elastic moduli and dispersion relations in colloidal glass, which recently could be measured by video microscopy in two dimensions. It is hoped that the review will motivate studies in three dimensions.  

\section{Introduction to elasticity in the context of thermodynamics}

The free energy of a crystal at constant temperature ($dT=0$) obeys the Gibbs fundamental form,  the  first law of thermodynamics connecting infinitesimal energy changes
\begin{equation}\label{firstlawF}
dF = -pdV +\mu dN + h_{\alpha\beta}dU_{\alpha\beta}\, .
\end{equation}
Besides the familiar mechanical and chemical terms, it includes a term with a stress tensor $h_{\alpha\beta}$ at constant volume $V$ and particle number $N$ times an extensive strain tensor $U_{\alpha\beta}=V u_{\alpha\beta}$. The corresponding work done is $\delta W = \int h_{\alpha\beta}\delta u_{\alpha\beta} dV$, 
 and $u_{\alpha\beta}=\frac12\left[\nabla_\alpha u_\beta({\bf r})+\nabla_\beta u_\alpha({\bf r})\right]$ is the symmetrized strain tensor field. The displacement field $\ub(\rb)$ describes the coarse-grained displacement of the particles from their unstrained sites; its proper definition will be given below. Because we remain close to unstrained equilibrium, the linearized form for $u_{\alpha\beta}$ suffices in the present context. Einstein's sum convention of repeated (Greek) indices  summed over  all spatial directions  is employed throughout the text. 

The elastic constants are defined as second derivatives of the free energy, which shall be taken at constant density
\begin{equation}\label{def}
\frac{\partial^2 F/V}{\partial u_{\alpha\beta} \partial u_{\gamma\delta}}\Big|_{n} = \frac{\partial h_{\alpha\beta}}{\partial u_{\gamma\delta}}\Big|_{n} = C^n_{\alpha\beta\gamma\delta}.
\end{equation}
Because of this definition, the tensor of fourth rank obeys the 
Voigt symmetry relations, $C^n_{\alpha\beta\gamma\delta}=C^n_{\beta\alpha\gamma\delta}=C^n_{\alpha\beta\delta\gamma}=
C^n_{\gamma\delta\alpha\beta}$, which follow as the strain tensor is symmetric, and the $C$ are second order derivatives. 

The above free energy is formulated to capture the existence of long-ranged correlations in the displacement field, which can be rigorously  proven using a Bogoliubov inequality \cite{Wagner1966} (see Sect.~3.3 \& 4.1) or motivated via the following heuristic argument within  linear response: The change in the macroscopic distortion $\nabla \ub$ in linear  order in a small applied homogeneous field $h_{\alpha\beta}$, is given by the thermodynamic derivative (susceptibility) $\chi$, $\chi = \delta \langle \nabla u \rangle/\delta h|_{h=0} $, which is according to the fluctuation-dissipation theorem  given by (brackets $\langle\ldots\rangle$ denote canonical averaging):
\begin{eqnarray}
k_BT {\bf \chi }&=& 
 \frac{1}{V}\!\!  \int\!\!\!\!\! \int\!\! d\rb d\rb' \;\langle  \nabla_\alpha \ub(\rb)\;  \nabla_\beta' \ub(\rb') \rangle
= -\frac{1}{3}  \int\!\! d\rb \; \nabla^2 \langle \ub(\rb) \;\ub({\bf 0}) \rangle ,\nonumber
\end{eqnarray}
when spatial homogeneity and isotropy are assumed at long distances for the sake of simplicity. Continuing in this simplified consideration, and assuming that the remaining integral is given by a rapidly varying function, one arrives at
\begin{eqnarray}
    \nabla^2 \langle \ub(\rb) \;\ub(\rb') \rangle = - \frac{k_BT}{R} \; \delta(\rb-\rb')\; ,\quad \Leftrightarrow \quad  \langle \ub(\rb)\; \ub(\rb') \rangle = \frac{k_BT}{4\pi R}\; \frac1{|\rb-\rb'| }\;,\nonumber
\end{eqnarray}
where $R\propto1/\chi$ is some measure of the elastic constant of the material. This heuristic argument for long-ranged correlations (for an isotropic solid it could easily be made rigorous solving the equations of linear elasticity theory \cite{Landau})  is backed-up by a proper and general  discussion of the correlations of the displacement field. They are given by the (inverse) of the dynamical matrix ${\cal D}_{\alpha\beta}(\mathbf{q})$ at wavevector $\qb$:
\begin{equation}\label{disprel}
\langle \delta u_\alpha^\ast(\qb) \;\delta u_\beta(\qb) \rangle = Vk_BT {\cal D}_{\alpha\beta}^{-1}(\mathbf{q})
\quad \mbox{and }\;
{\cal D}_{\alpha\gamma}(\mathbf{q})\to C^c_{\alpha\beta\gamma\delta}\; q_\beta q_\delta\quad\mbox{for } q\to0.
\end{equation}
The small-$q$ divergence like $1/q^2$ corresponds to the $1/r$ long-ranged nature in real space.
Note the subtle difference that $C^c_{\alpha\beta\gamma\delta}$ in Eq.~\eqref{disprel} is taken at fixed defect concentration $c$, while $C^n_{\alpha\beta\gamma\delta}$ in Eq.~\eqref{def} is at constant density $n$; an example of the (sometimes cumbersome) considerations required in  disordered solids; see Sect.~3.3 and the Appendix.
The eigenvalues $\lambda_s(\qb)$ (the subscript $s$ denotes polarization) of the dynamical matrix ${\cal D}(\qb)$ can be considered the spring constants of the system, and are called 'dispersion relations'. In atomistic solids, they determine the eigenfrequencies of lattice vibrations, but in colloidal solids the motion is over-damped due to the surrounding solvent. Nevertheless, the dispersion relations characterize the elastic energy stored in plane wave distortions (with wavelength $2\pi/q$) of the solid. 

\section{Crystalline solids}

This section focuses on microscopic expressions for the tensor of elastic coefficients and the dispersion relations in real, mono-domain crystals, pointing to the corrections from local disorder or non-affine deformations. Because of the advances in microscopy of colloidal solids, the non-ideality of samples obtained by self-assembly plays an important role in the analysis of data and can also be studied in detail. Theoretical concepts incorporating disorder will be presented, even though specific results on defect diffusion and defect-strain coupling will not be discussed  for lack of space. Those aspects can be found in the literature which will be mentioned.  

\subsection{Born term}

The reference result for the calculation of elastic constants from the particle interactions was obtained by Born \cite{born}, and follows from an expansion of the potential energy for small displacements of the particle positions around their equilibrium sites ($i=1,\ldots,N$ counts through the particles)
\begin{equation}
C^B_{\alpha\beta\gamma\delta} = \frac{1}{2V}\sum_{i,j}
\Phi_{\alpha\gamma}(i,j) R_{ij,\beta}R_{ij,\delta}\;
,\label{A2limit}
\end{equation}
with $\Phi_{\alpha\gamma}(i,j) =
\nabla_\alpha^{(1)}\nabla_\gamma^{(2)} \Phi(\Rb_{ij})$,
and the actual pair potential $\Phi$. The derivatives are evaluated at the
equilibrium positions $\Rb_i$, and $\Rb_{ij}=\Rb_i-\Rb_j$. While in ideal crystals  which is the situation originally considered periodic lattice sites were used to define  $\Rb_i$, in later application to colloidal solids the time averaged centers of the particle trajectories were employed
\cite{Keim2004,Reinke2007}.  Regions without point defects were studied, because defect diffusion would lead to large excursions of single particles from their lattice sites, thereby invalidating the expansion; see Sect.~3.4. They also would destroy the regularity  of the real space lattice, so that the dispersion relations would not be periodic from one Brillouin zone to another.

\subsection{Fluctuation term}

The above  Born term neglects thermal fluctuations, as only the change in potential energy is considered. At finite temperatures the free energy needs to be considered and its variation with strain. Motivated by the need to analyze their atomistic computer simulations, Squire, Holt, and Hoover derived expressions for the isothermal elastic constants valid at finite $T$ and for standard particle pair potentials  \cite{Squire1969}. The relations between different expressions were further clarified \cite{Lutsko1989}, and the formula combining  Born and the so-called  fluctuation term nowadays are the basis for computer simulations studies of elasticity in solids, including  disordered ones  
\cite{Barrat2006}. In principle, they could also be evaluated by microscopy in colloidal solids if the pair potential is smooth and well known. 

\begin{figure}[!h]
\includegraphics[width=0.45\textwidth]{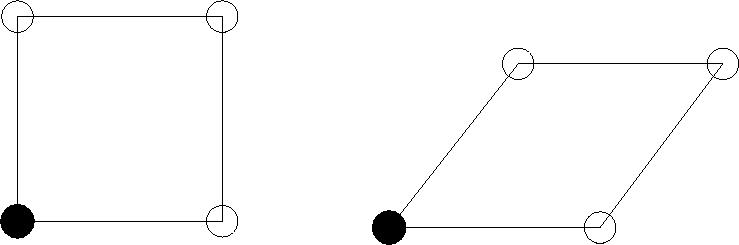}\hspace*{.5cm} \includegraphics[width=0.45\textwidth]{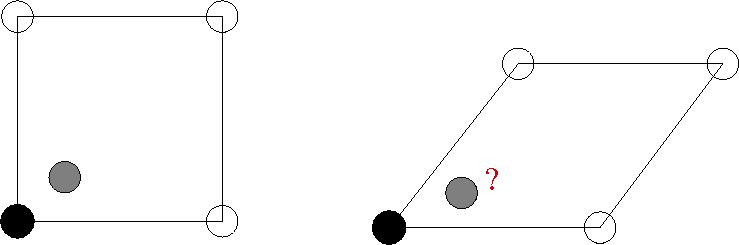}     
\caption{The left two panels show an affine shear deformation of a unit cell of a crystal, where all particles move according to the  macroscopically enforced distortion. The right two panels include an interstitial particle (or the second particle in a basis), whose displacement after the affine deformation remains open. Non-affine local particle displacements are especially important in crystals with complicated unit cells or in disordered solids. }
\end{figure}

Figure 1 presents the conceptual problem when determining elastic energies from macroscopically imposed deformations, especially in crystals with complicated unit cells or in disordered systems. The imposed macroscopic deformation does not affinely determine the individual particle displacement as assumed in the derivation of the Born term. Particles can be displaced  non-affinely, which even at vanishing temperature causes a correction to the Born-term in the elastic constants \cite{Lutsko1989,Barrat2006}. The nature of the displacement of the particles needs to be determined and enters as constraint on the variation  when calculating elastic constants. In a deformation, the particles move to positions where the forces cancel. When considered at vanishing temperature, this condition requires a 'relaxation' of the particles to the adjacent (potential) energy minima. In the canonical ensemble, where the average energy is given, the local stresses also perform no work on average, which states that the constraint of vanishing forces also holds at finite temperatures.  Consider the free energy $F=-k_BT \ln Z$, which is given by the  partition sum of the canonical ensemble $Z=\int d{\bf p}^{N} d\rb^{N} \ e^{-\beta \hat{H}}$,  where the Hamiltonian reads  $\hat H=\sum_i\frac{{\bf p}_i^2}{2m}+\frac12\sum_{i\ne j}\Phi(\rb_{ij})$. The stress field $\bf t$ is the field thermodynamically conjugate to strain 
\begin{equation}
t_{\alpha\beta}= \langle \hat{t}_{\alpha\beta}\rangle=\frac 1V \frac{\partial F}{\partial u_{\alpha\beta}}=\frac{-k_BT}{VZ}\frac{\partial Z}{\partial u_{\alpha\beta}}\label{eq5}\;.
\end{equation}
For performing the derivative, an affine mapping of arbitrary points in the material is applied to the Hamiltonian function, the linear change in $Z$ obtained (note that $V$ is varied in this procedure), and the mapping is inverted to obtain the corresponding (macroscopic) strain $u_{\alpha\beta}$ entering thermodynamics. This result for the average (Cauchy) stress can be obtained from averaging the Irving-Kirkwood expression of the microscopic stress tensor
\begin{equation} \hat{t}_{\alpha\beta}= \frac{-1}{V} \left[ \sum_{i=1}^N  \frac{p^i_\alpha p^i_\beta}{m} -\frac 12 \sum_{i\ne j}^N \left(
\frac{ r_{ij,\alpha} r_{ij,\beta}}{r_{ij}} \frac{\partial \Phi(r_{ij})}{\partial r_{ij}} \right) \right]\ .\label{kirkwood}
\end{equation}
The first contribution comes from the kinetic energy, the second from the virial and the potential energy. The microscopic stress tensor $\bf \hat t$ also determines the conservation law $\partial_t \hat{j}_\alpha + \nabla_\beta\hat{t}_{\alpha\beta}=0 $ of the momentum density $\hat{\bf j}(\rb,t)$.  In unstrained thermal equilibrium, the stress tensor averages to the negative of the pressure (see the Appendix): 
\begin{equation}\label{pressure}
\left.t_{\alpha\beta}\right|_{u_{\gamma\delta}=0} =- p\; \delta_{\alpha\beta} \quad\mbox{unstrained.}
\end{equation}
The elastic constants at vanishing external strain  follow as second  derivatives, repeating the above procedure, and contain three terms 
\begin{eqnletter}\label{fluctuationterm}
C_{\alpha\beta\gamma\delta}^{(t)} &=&  \frac{\partial t_{\alpha\beta}}{\partial u_{\gamma\delta}}= \frac{\partial }{\partial u_{\gamma\delta}} \Big(\frac{1}{{Z}}\int dp^{3N} dq^{3N} { \hat{t}_{\alpha\beta} } e^{-\beta{\hat{H}}}\Big)\nonumber \\
 &=& \quad {2nk_BT(\delta_{\alpha\gamma}\delta_{\beta\delta}+\delta_{\alpha\delta}\delta_{\beta\gamma})}\\
 & & +\;  { \langle C^B_{\alpha\beta\gamma\delta}\rangle} \\
 & & -\; \frac{1}{k_BT} \Big[{\langle \hat{t}_{\alpha\beta}\hat{t}_{\gamma\delta}\rangle} - {\langle \hat{t}_{\alpha\beta}\rangle\langle\hat{t}_{\gamma\delta}\rangle}\Big]
\end{eqnletter}
The term (\ref{fluctuationterm}a) is kinetic in origin, while the canonically averaged Born term (\ref{fluctuationterm}b) follows from the derivative acting on  $\bf \hat{t}$. The variance of the Irving-Kirkwood stress tensor, the first term in (\ref{fluctuationterm}c) follows from the variation of the Hamiltonian function in the average, while the square of $\langle\hat{t}_{\gamma\delta}\rangle$ originates from the normalization $1/Z$. In the Appendix the relation is shown from Eq.~\eqref{fluctuationterm} to the elastic constants at fixed, possibly vanishing, defect concentration.  Because no mapping of the particle positions to periodic lattice sites is involved, the above expression has turned out useful for the study of glasses and other disordered materials especially at low temperatures; for a more in depth discussion and results on glasses see the review \cite{Barrat2006}.

\subsection{Density functional approach}

Density functional theory (DFT) is especially powerful in describing inhomogeneous states like crystalline solids. There exists an extensive body of work, using DFT to obtain the elastic constants of periodically ordered solids, which is partially  reviewed in Ref.~\cite{Kirkpatrick1990,Masters2001,Walz2010}. Yet application of DFT to this problem requires an ad hoc ansatz of  the inhomogeneous density field as expressed by the coarse-grained fields of elasticity theory. Different choices have been made in order to incorporate defect density, because no fundamental criterion had been known to motivate the ansatz a priori. Also the ansatz was extended to glasses \cite{Masters2001}, which appears questionable in view of the material reported in Sect.~4.3. The recent developments on elasticity in non-ideal cystalline solids overcame this problem by going back to the microscopic inhomogeneous density field and studying its time dependence  \cite{Walz2010}. As will be discussed in Sect.~3.4, at present the most stringent connection between different approaches to elastic coefficients appears to be via the time-dependent equations of motion. Before entering into a discussion of these finer points concerning  DFT, here the elastic free energy shall be derived from DFT following the standard steps in order to familiarize the reader with the approach \cite{Walz2012}. DFT expressions are especially suited to hard spheres, which are the basis of most colloid-particle models,  because the explicit potential and its derivative [in the Virial of Eq.~\eqref{fluctuationterm}] are circumvented; these  would   not be well defined.

To obtain the second order derivatives of the free energy with respect to its thermodynamic variables, one starts from the second order change in free energy $\Delta \mathcal{F}$ due to a variation in the density distribution $\delta \rho(\rb)$ around the periodic crystalline equilibrium density \cite{rowlinson}:
\begin{equation}\label{dft}
\beta \Delta \mathcal{F} = \frac{1}{2}\int\!\!\int d^d\!r_1 d^d\!r_2 \Big[\frac{\delta(\mathbf{r_{12}})}{n(\mathbf{r_1})}-c(\mathbf{r_1,r_2})\Big] \delta \rho(\mathbf{r_1})\delta\rho(\mathbf{r_2}),
\end{equation}
where $c(\mathbf{r_1,r_2})$ is the direct correlation function of a periodic crystal.  The average periodic density is $n(\rb)$, which can be observed in e.g.~Bragg scattering. The $\delta \rho(\rb)$ is the variation in the microscopic density profile, which varies on length scales extending down to e.g.~particle size and lattice constant.   Assuming that the integral is dominated by a smooth variation described with the ansatz
\begin{equation}\label{ansatz}
\delta\rho(\rb) = -\mathbf{u(r)\cdot\nabla}n(\rb) \, ,
\end{equation}
one can simplify Eq.~\eqref{dft} in order to evaluate the thermodynamic derivatives appearing in Eq.~\eqref{def}.  Here ${\ub}({\rb})$  is the coarse-grained displacement field and its symmetrized gradient is the macroscopic strain field. The ansatz for the density change is motivated by a macroscopic consideration where the smoothly strained average density profile $n\left(\rb+\ub(\rb)\right)$ is expanded in linear order in $\ub$. In  the ansatz~\eqref{ansatz}, the fields live on two very different length scales. The coarse-grained displacement field describes long wavelength fluctuations arising in elasticity theory. The local density profile varies rapidly from one lattice plane to the next, as does the direct correlation function which is the second functional derivative of the (interaction) free energy functional  in DFT. Both locally varying functions are connected by an important relation in DFT, which is the crucial ingredient in deriving the proper thermodynamic derivatives from the above $ \Delta \mathcal{F}$.
The integrals can be manipulated with the help of the Lovett, Mou, Buff, and Wertheim (LMBW) equation
\begin{equation}\label{LMBW1}
\frac{\nabla_\alpha n(\rb)}{n(\rb)} = \int d^d\!r^\prime \; c(\mathbf{r,r^\prime})\;  \nabla^\prime_\alpha\, n(\mathbf{r^\prime})\;.
\end{equation}
This equation is a statement of the translational invariance of the underlying Hamiltonian function of the system, which is spontaneously broken in the crystal state  but constrains the direct correlation function and the average density profile \cite{rowlinson,lmbw1,lmbw2}.  To employ Eq.~\eqref{LMBW1},
one substitutes Eq.~\eqref{ansatz} into  $ \Delta \mathcal{F}$ and expands  one slowly varying displacement field $u_\beta(\mathbf{r_2})$ around $\mathbf{r_1}$, the position of the other. The zero order term of $\Delta\mathcal{F}$ vanishes because of   Eq.~\eqref{LMBW1} and the first order term does not contribute as $c(\mathbf{r_1,r_2})=c(\mathbf{r_2,r_1})$. Truncating the expansion at the squared gradient level, because  the hydrodynamic variable $\mathbf{u(r)}$ is assumed slowly varying, one obtains an expression which is quadratic in $\nabla\mathbf{u(r)}$
\begin{eqnarray}
\Delta\mathcal{F} &=& \frac{k_BT}{2}\int\!\! \int d^d\!r_1 d^d\!r_2 \Big[\frac{\delta(\mathbf{r_{12}})}{n(\mathbf{r_1})}-c(\mathbf{r_1,r_2})\Big] u_\alpha(\mathbf{r_1})u_\beta(\mathbf{r_2}) \nabla_\alpha n(\mathbf{r_1})\nabla_\beta n(\mathbf{r_2})\nonumber\\
 &=& \frac{k_BT}{2}\!\!\int\!\!\! \int\! d^d\!r_1 d^d\!r_2 \nabla_\alpha n(\mathbf{r_1})c(\mathbf{r_1,r_2})\nabla_\beta n(\mathbf{r_2}) u_\alpha(\mathbf{r_1})
\nonumber\\ &  &\Big[u_\beta(\mathbf{r_1})\!-\!u_\beta(\mathbf{r_1})\!+\nabla_\gamma u_\beta(\mathbf{r_1}) r_{12,\gamma}- \frac{1}{2}\nabla_\gamma\nabla_\delta u_\beta(\mathbf{r_1}) r_{12,\gamma}r_{12,\delta}+\ldots\Big]\nonumber \\
 &=& \frac{1}{2}\int d^d\!r \; \lambda_{\alpha\beta\gamma\delta} \nabla_\gamma u_\alpha(\rb) \nabla_\delta u_\beta(\rb),\label{Cresult1}
\end{eqnarray}
with expansion coefficient tensor
\begin{equation}\label{Cresult}
\lambda_{\alpha\beta\gamma\delta} = \frac{k_BT}{2V}\int\!\!\int d^d r_1 d^d r_2 \;\left[ \nabla_\alpha n(\mathbf{r_1})\right]\; c(\mathbf{r_1,r_2})\;\left[\nabla_\beta n(\mathbf{r_2})\right]\; r_{12,\gamma}\;r_{12,\delta}
\end{equation}
To derive the required   Voigt symmetry of the elastic constants, one requires the rotational analog of the LMBW equation as explained in Ref.~\cite{Walz2010}. It arises from the isotropy of the underlying Hamiltonian and leads to (in $d=3$ dimensions) 21 independent $\lambda$-elements, which obey the relations: $\lambda_{\alpha\beta\gamma\delta}=\lambda_{\beta\alpha\gamma\delta}=\lambda_{\alpha\beta\delta\gamma}=
\lambda_{\gamma\delta\alpha\beta}$. Then the gradients of the displacement field can be symmetrized to get the strain fields and the four equivalent versions of Eq.~\eqref{Cresult1} can be combined. Exploiting the symmetries of $\lambda$ further, the expressions for the elastic coefficients can be condensed  \cite{wallace}. Finally, the derivatives in  Eq.~\eqref{def} can be evaluated for  homogeneous strain to find:
\begin{equation}\label{Cresult2}
C^n_{\alpha\beta\gamma\delta}=\lambda_{\alpha\gamma\beta\delta}+ \lambda_{\beta\gamma\alpha\delta}-\lambda_{\alpha\beta\gamma\delta}\;.
\end{equation}
The $C^n$ inherit the Voigt symmetry from the $\lambda$ even though different index combinations appear.
To evaluate the DFT result for the elastic constants,  an approximation of the direct correlation function of the crystal and for the average density profile is required. Both could be taken e.g.~from the Ramakrishnan-Yussouff approach to hard sphere crystals based on the Percus-Yevick approximation \cite{Chaikin95}. 

The above result neglects the coupling between strain and density fluctuations,  and has been generalized to include defect densities in various ad hoc ways \cite{Walz2010,Kirkpatrick1990,Masters2001}. In order to find the proper way, it is advantageous
to combine DFT with the Zwanzig-Mori  (ZM) equations of motion of the microscopic density fluctuations \cite{Forster,Chaikin95}.  Thereto, only the reversible couplings are required,  which can be formulated without approximations. The criterion for the selection of the variables in the ZM approach is given by the Bogoliubov-inequality see [Eq.~\eqref{bogol}], which yields all long-ranged correlations in a crystal \cite{Wagner1966}. This enables one to generalize the ansatz~\eqref{ansatz}  to include the defect density, $\delta c(\rb) $ \cite{Walz2010}:
\begin{equation}\label{ansatz2}
\delta\rho(\rb) = -\nabla\cdot \left[  n(\rb) \mathbf{u(r)}  \right]- \frac{n(\rb)}{n_0} \delta c(\rb) \, ,
\end{equation}
Again, the coarse-grained fields  $\mathbf{u(r)}$ and $\delta c(\rb)$ are assumed to vary much more smoothly than the average crystalline density profile $n(\rb)$. The first term, which is the only one present in an ideal crystal, now properly accounts for the strain-density coupling: the shape fluctuation of Eq.~\eqref{ansatz} is complemented by a volume term so that the density change is given by a divergence term \cite{Kirkpatrick1990}; this follows from mass conservation as lattice motion is identical to particle motion in an ideal crystal. The second term in Eq.~\eqref{ansatz2} then couples in the independent defect motion.
The ansatz \eqref{ansatz2} that the slow contributions in the microscopic density fluctuations $\delta\rho(\rb,t)$  are dominated by four coarse-grained fields  $\delta c(\rb) $  and $\ub(\rb) $  solves the reversible, isothermal ZM equations for the microscopic density fluctuations $\delta\rho(\rb,t)$  on all length scales. While its uniqueness has not yet been shown, it appears unlikely that the DFT approach could lead to another relation.  Another advantage of the approach is the possibility to invert the relation and obtain microscopic definitions of the fields of elasticity theory from the microscopic density fields; see Sect.~3.4.
The above coarse-graining can  be repeated including all elastic fields of the free energy density $f(u_{\alpha\beta},c)=F/V$. The $f$ is the starting point for deriving hydrodynamic equations in the framework of non-equilibrium thermodynamics \cite{Martin1972}, and (for completeness) is derived from the free energy Eq.~\eqref{firstlawF} in the Appendix. Its term $\Delta f$ quadratic in the thermodynamic fields  contains all constants of elasticity of a general crystal,  $\nu$, $\mu_{\alpha\beta}$, and $C_{\alpha\beta\gamma\delta}$, in the general case; see  Ref.~\cite{Walz2012} for definitions and  calculation: 
\begin{eqnarray}\label{FreeEnC}
 \Delta f  &=&\!\frac{k_BT}{2}  \left\{\;  \frac{\nu}{n_0^2}\;\delta c^2
 + 2 \frac{ \nu \delta_{\alpha\beta} +\mu_{\alpha\beta}}{n_0}\, \delta c\, u_{\alpha\beta}
\right. \nonumber\\ & & + \left. 
\left[\;C^n_{\alpha\beta\gamma\delta}\!+\!\nu\delta_{\alpha\beta}\delta_{\gamma\delta}\!+\!\mu_{\alpha\beta}\delta_{\gamma\delta}\!+\!\delta_{\alpha\beta}\mu_{\gamma\delta}\; \right]\; u_{\alpha\beta}\, u_{\gamma\delta}\; \right.\Big\}.
\end{eqnarray}
Because of the coupling between density and the trace of the strain field, there is an effect on the elastic constants. The angular bracket $[\ldots]$ defines $C^c$, the elastic coefficients at constant defect density, while $C^n$ is at constant density. 

\subsection{On the comparison  of the different approaches}

\begin{figure}[!h]
\begin{center}
\includegraphics[width=0.3\columnwidth]{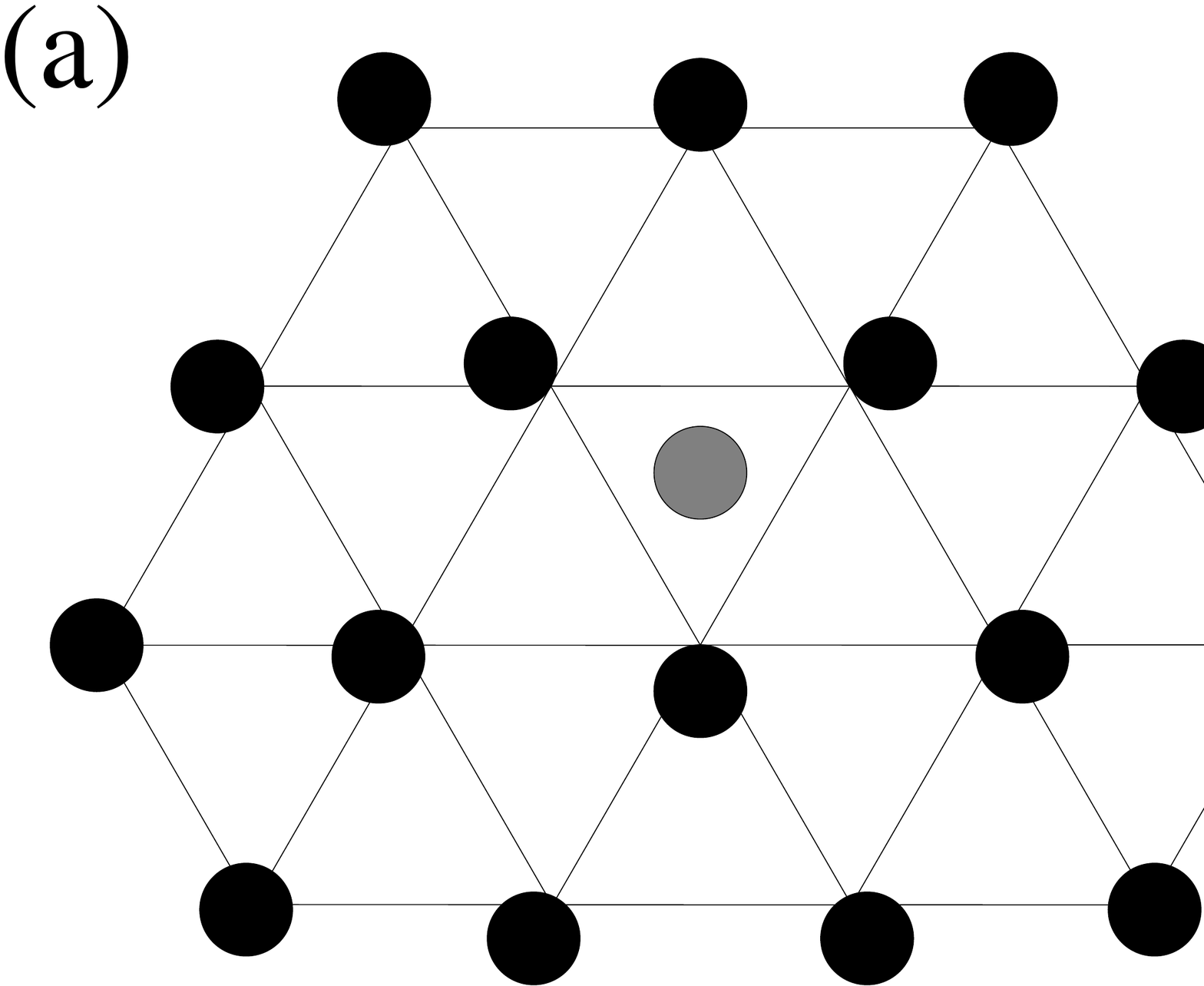}\hspace*{.5cm}
\includegraphics[width=0.3\columnwidth]{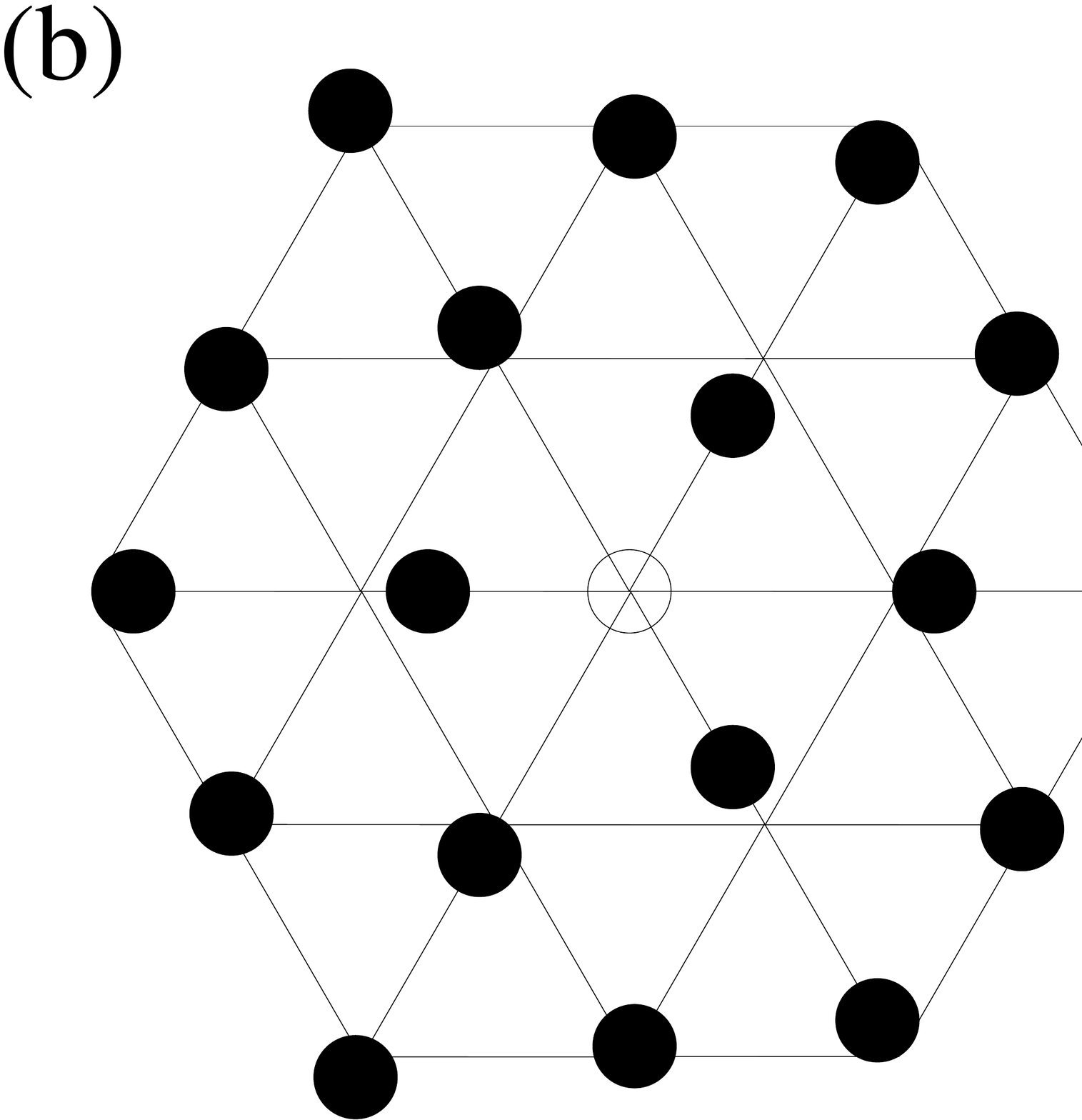}\hspace*{.5cm}
\includegraphics[width=0.2\textwidth]{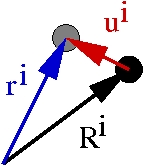} 

\vspace*{-4cm}
\rightline{\bf \large (c) \hspace*{2cm}}
\vspace*{4cm}

\caption{ Schematic two dimensional drawing of a crystal containing point defects, in panel (a) an interstitial and in panel (b) a vacancy; from Ref.~\cite{Walz2010}. In panel (c), the definition of the displacement vector $\ub_i$ of particle $i$ is given as difference between actual particle position $\rb_i$ and reference site $\Rb_i$. In case of point defects, particles can not be labeled by the lattice positions, as implied when evaluating the Born term \eqref{A2limit}  as sum over lattice sites. }
\end{center}
\end{figure}

In ideal solids where each site in a periodic lattice is occupied precisely by one particle, defining displacement and strain fields and the corresponding elastic constants is straightforward. With point defects, like vacancies or interstitials, defining the displacement field becomes more complicated; see Fig.~2. Also the non-affinity of the local particle displacements becomes more important \cite{Barrat2006}. While in equilibrium hard sphere crystals, the defect density of vacancies or  interstitials is low at melting ($c_v=10^{-4}$ and $c_i=10^{-8}$, respectively), it increases dramatically when the particles have a polydisperse size distribution as is always present in colloidal solutions; then the density of interstitials can reach $c_i=0.02$ before the crystal melts for even higher polydispersity \cite{Pronk2004}. Then the coupling of defect density to strain fluctuations can not be ignored anymore, and the terms $\mu$ and $\nu$ in Eq.~\eqref{FreeEnC} matter. Thus it is of interest to compare the different expressions for $C_{\alpha\beta\gamma\delta}$, to ascertain in which situation the correct result is obtained. 

A first observation concerns  the Born term \eqref{A2limit}, which can be recovered from the DFT result [Eq.~\eqref{Cresult}] using the so-called random phase approximation (RPA) 
$c(\mathbf{r_1,r_2})=-\beta \Phi(\mathbf{r_1,r_2})$ and
appropriately coarse-graining. Reassuringly, this most simple DFT approximation to replace the direct correlation function with the true pair potential  \cite{rowlinson}, recovers the approximation following from affine particle displacements at vanishing temperature. 

The 
wave equation and the complete equations of motion of elasticity theory provide at present the crucial test for the derived relations of the coarse-grained  fields and their elastic coupling coefficients. 
Considering only the wave equation for the momentum field, this requires to establish 
\begin{equation}
mn_0\partial^2_t \delta j_\alpha(\mathbf{q},t)  = - {\cal D}_{\alpha\beta}(\mathbf{q})  \delta j_\beta (\mathbf{q},t)\;,  
\end{equation}
with the dynamical matrix ${\cal D}_{\alpha\beta}(\mathbf{q}) $ defined in Eq.~\eqref{disprel} from the long-ranged displacement fluctuations via the equipartition theorem. The momentum density ${\bf j}(\mathbf{q},t)$ is the familiar one, also used in hydrodynamics. From the phenomenological approach, the wave equation and actually the complete long-wavelength coupling including defect density is well established and contains a number of subtle relations like the Voigt symmetries \cite{Martin1972,Fleming1976}. Any microscopic approach needs to recover this. In the DFT approach, the ZM equations show that $C^c$ and not $C^n$ determines  the sound velocities, viz.~the long-wavelength limit of the wave equation \cite{Walz2010}.  Beyond this, the dispersion relations for finite $\qb$ can be compared.
This route to establish microscopic expressions of elastic constants appearing in states with spontaneous symmetry breaking is reminiscent of the comparison of the  Triezenberg-Zwanzig and the 
Kirkwood-Buff expression of the surface tension. These two relations, one containing the direct correlation function and the other the actual pair potential,  can also be identified via the dynamic equations of motion  \cite{rowlinson}. Reassuringly in the present consideration, the DFT approach leads to the complete reversible and isothermal equations of crystal elasticity with all reversible couplings and their symmetries derived \cite{Fleming1976}; the generalization to include heat transport and dissipation are presently investigated.
It would be desirable to establish rigorously that the expression \eqref{fluctuationterm} including the fluctuation term agrees with the DFT expression \eqref{FreeEnC} containing the defect corrections. But while we expect this to hold, the coupling of defects has not been addressed explicitly in the former approach.   
\begin{figure}[!ht]
\begin{center}
\includegraphics[width=0.45\textwidth]{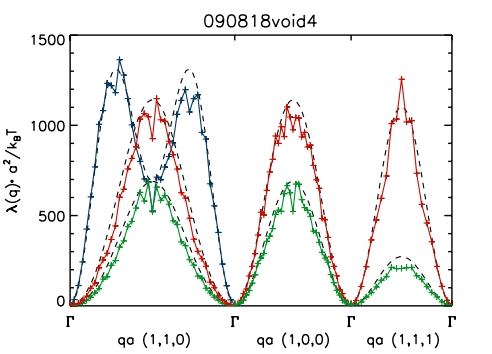}\hspace*{.5cm} 
\includegraphics[width=0.45\textwidth]{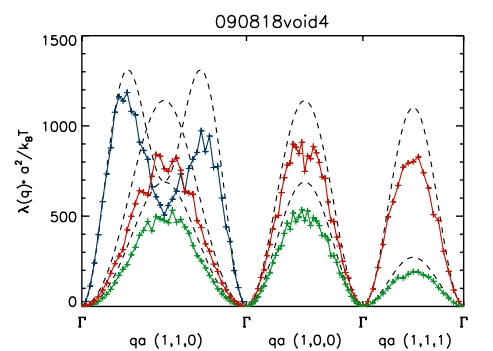}\\
\includegraphics[width=0.45\textwidth]{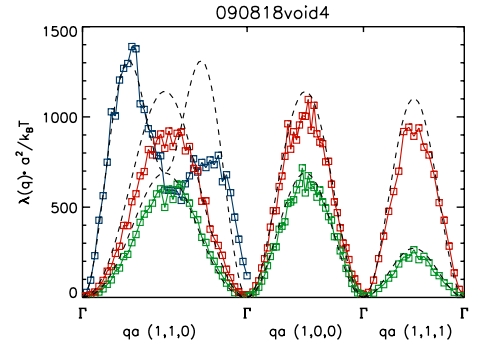}\hspace*{.5cm} 
\includegraphics[width=0.45\textwidth]{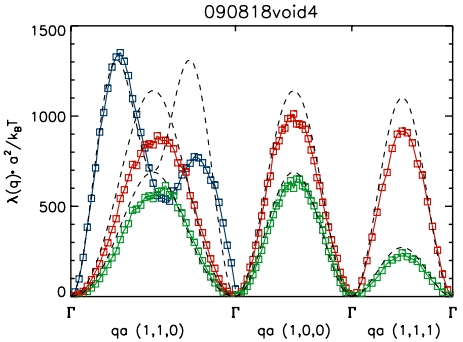}
\end{center}
\caption{Dispersion relations of a defective colloidal hard sphere like crystal (of fcc structure)  for various high symmetry directions $\qb$  in reciprocal space ($\qb$ given below the x-axis,  starting from $\qb=0$, the $\Gamma$ point, and symmetrically extended into the 2nd Brillouin zone); colors give different polarizations. The results are obtained from MC simulations of the variance of displacements according to  Eq.~\eqref{disprel} \cite{urs}. The upper row uses time averaged lattice sites to calculate particle displacements, the lower row uses the DFT expression for the displacement field obtained from density fluctuations. Left panels average displacements over 100 MC steps, right panels over 500. While the DFT dispersion relations improve with better statistical averaging, the upper row results neglect defect motion and deteriorate with averaging time. Dashed lines give $T=0$ calculations using  the pair potential based on Eq.~\eqref{A2limit} evaluated at periodic lattice sites.}
\end{figure}

To give an example of the comparison of different approaches,
Fig.~3 shows the effects of local defects on the dispersion relations obtained in computer simulations \cite{urs}. The eigenvalues of  ${\cal D}_{\alpha\beta}(\mathbf{q})$ as functions of wavevector, are presented for a hard sphere crystal with only few defects. To be precise, Monte Carlo (MC) simulations of 32003  hard sphere Yukawa particles were performed mimicking the colloidal  system of Ref.~\cite{Reinke2007}, and 4 vacancies and 7 interstitials were introduced by hand. Equilibrating the defects for $7\times 10^4$ MC steps and then measuring the displacement fluctuations over a time span $\Delta t$ of 100 or 500 MC steps,  the dispersion relations were obtained from Eq.~\eqref{disprel}. Figure 3 shows that defect motion affects the measurement of the elastic relations depending on the definition of displacement fluctuations used. Even a tiny number of defects can destroy the accuracy of the dispersion relations if the defects are mobile and the measurement averages over a finite time; this happens in the upper row of Fig.~3. It holds in the method defining particle displacements as difference vectors of the particles momentary position from the time-averaged center of the trajectory; see Fig.~2(c) and Eq.~\eqref{eq0} below for the precise definition. This method  thus neither requires a periodic lattice nor fails in assuming an one-to-one mapping of particles to (average) sites.  Yet it is incorrect and fails because after extending the averaging time span $\Delta t$, particles hop on vacancy positions and move arbitrarily far following the diffusion of the defects.  A low but finite density of particles appears to be displaced far wider than corresponding to the correct local elasticity. In the long run, the variance of  displacements of a low but finite density (closely related to the  defect density) of particles will become arbitrarily large, and thus the apparent dispersion relations will be incorrectly low. Note the decrease in the so measured dispersion relations in the upper right side panel of Fig.~3. Dashed lines are $T=0$ calculations based on the potential evaluated at the lattice sites $\Rb_i$. Because of the periodicity of the $\Rb_i$, the theoretically calculated dispersion relations are periodic in reciprocal space; note that this does not hold in the simulated ones because the time-averaged centers of the particle trajectories do not lie on a perfect lattice. This however should not be considered an error of the simulation method, because the  one-to-one mapping of particles to lattice positions $\Rb_i$ is impossible in the presence of point defects. The perfect lattice and the periodicity of the dispersion relations are an idealization only possible in an ideal crystal.  

The lower panels in Fig.~3 were obtained with displacement fluctuations calculated via the inverted relation \eqref{ansatz2} in the DFT approach. It can best be inverted in Fourier space and expresses the coarse-grained displacement in terms of microscopic density fluctuations, e.g.~obtainable from a time-dependent DFT approach  \cite{Walz2010,Szamel1993}:
\begin{equation}\label{displacement}
\ub(\qb,t) = i {\bf {\cal  N}}^{-1} \sum_{\gb} n_{\gb}^\ast\, {\gb} \; \delta \rho_{\gb}(\qb,t)\;,
\end{equation}
with a normalization matrix ${\bf {\cal  N}} = \sum_\gb |n_\gb|^2 {\gb}{\gb}$. 
Only spatially modulated density fluctuations $\delta \rho_{\gb}(\qb,t)=\int\!\!d^dr\;e^{-i(\gb+\qb)\cdot \rb}\; \rho(\rb,t)-n_\gb V \delta_{\qb,0}$ need to be measured, which require no definition of a lattice in real space. The crystal structure enters in Fourier space via the reciprocal lattice vectors $\gb$ and the Bragg-scattering amplitudes $n_\gb$; the latter are the order parameters of the crystal, and describe the modulated average density $n(\rb)=\sum_{\gb} \; e^{-i\gb\cdot\rb}\; n_\gb$. Local defects lower the magnitude of the   $n_\gb$ (increase the Debye-Waller factor, like thermal fluctuations also do), but do not destroy the reciprocal lattice \cite{Chaikin95}. Because the thus defined displacement is a continuous field, its dispersion relations are not periodic in reciprocal space; this can be  seen in Fig.~3 and should not be taken as a defect of the definition.  Averaging longer over time increases the accuracy in the sampling of the variance of displacement fluctuations and thus reduces the noise in the curves; this can be seen from comparing the lower right panel with the lower left one. The temperature in the simulation is quite low, so that the $T=0$ calculation describes the dispersion relations quite well in the first Brillouin zone. Temperature dependence can be seen only close to the Brillouin zone boundary, e.g. at and beyond point $K$, which lies around $1/3$ of the $\Gamma$-$\Gamma$ distance along the $[1,1,0]$ direction  \cite{Reinke2007}. In Figs.~3 and 4, $a$ is the average particle distance. For application of the sums in Eq.~\eqref{displacement} to real experiments, it is good news that only 14 Bragg peaks were included in the results of Fig.~3.

Reassuringly, the general definition of the displacement field in Eq.~\eqref{displacement} reduces to the expected low temperature result in ideal crystals. Let particle $i$ fluctuate around its site $\Rb_i$ in a periodic lattice. Then up to linear order in the displacement $\ub_i(t)=\rb_i(t)-\Rb_i$ the density fluctuation becomes
\begin{eqnarray}
\delta \rho_{\gb}(\qb,t) = \sum_{i=1}^N \; e^{-i (\gb+\qb)\cdot\rb_i(t)}\approx  \sum_{i=1}^N \; e^{-i \qb \cdot\Rb_i}\; \left( 1 - i \gb \cdot \ub_i(t) + {\cal O}((\gb\cdot\ub_i)^2) \right)\; ,
\nonumber\end{eqnarray}
where $e^{-i\gb\cdot\Rb_i}=1$, and $q\ll g$ was used. Inserting this into  Eq.~\eqref{displacement}, and using that 
$\sum_{\gb} n_{\gb}^\ast\, {\gb}=0$ by inversion-symmetry, gives
\begin{eqnarray}
\ub^{\rm id.cryst.}(\qb,t) =   \sum_{i=1}^N \; e^{-i \qb \cdot\Rb_i}\; \left( \ub_i(t) + {\cal O}((\gb\cdot\ub_i)^2) \right)\;, 
\end{eqnarray}
where the normalization by ${\bf {\cal  N}}$ could be canceled. This is
the expected result for an ideal periodic crystal.  It is periodic in reciprocal space, $\ub^{\rm id.cryst.}(\qb+\gb,t)=\ub^{\rm id.cryst.}(\qb,t)$, which the original expression \eqref{displacement}  was not; the difference is hidden in the neglected terms of order ${\cal O}((\gb\cdot\ub_i)^2)$.

\section{Glasses}

This section focuses on microscopic expressions for the tensor of elastic coefficients and the dispersion relations in amorphous solids, especially in colloidal glass. Video microscopy measurements of a binary glass former in two dimensions will be presented and the theoretical understanding will be reviewed. Again the question on how to define a displacement field and to use the equipartition theorem Eq.~\eqref{disprel} will be central.  

\subsection{Modified density functional approach}

The first theory to be  considered is 
Masters'  DFT-based approach \cite{Masters2001} where the coarse-graining of the DFT free energy described in Sect.~3.3 is applied to glass. Reconsidering the developments leading from Eq.~\eqref{dft} to  Eq.~\eqref{Cresult}, all steps appear to go through in amorphous solids as well as in crystalline ones; also the equipartition theorem Eq.~\eqref{disprel} seems valid at first inspection in glass. Yet the expression \eqref{ansatz} for the coarse-grained density field implies a long-ranged pair correlation function, and thus a small-$q$ divergence of the static structure factor in glass   \cite{Masters2001}. This should come as no surprise, as Wagner showed that the structure factor in crystals obeys the Bogoliubov inequality \cite{Wagner1966,Walz2010}:
\begin{equation}
\left\langle |\delta\rho(\mathbf{{g}}+\mathbf{{q}})|^{2} \right\rangle \ge \frac{(\mathbf{{g}}+\mathbf{{q}})^{2}(k_{B}T)^{2} |n_{\mathbf{{g}}}|^{2}V}{R {q}^{2}}\propto {q}^{-2} \;.\label{bogol}
\end{equation}  
where $R$ stands for one of the elastic coefficients depending on direction of $\gb$.  This diffuse scattering background  diverges for wavevector $q\to0$, but the divergence has a non-vanishing amplitude only around a finite reciprocal lattice vector, $\gb\ne0$, where the Bragg-peak resides. (The latter arises from constructive interference and thus is an order in particle number $N$ larger than the diffuse background.)  The divergence is strongly related for all Bragg-peaks, at it is coupled to the long-ranged displacement correlations in Eq.~\eqref{disprel}. In the amorphous glass, the concept of the inhomogeneous average density field $n(\rb)$ whose gradient has to be non-vanishing in the ansatz \eqref{ansatz} and which also appears in the long-ranged pair correlation function $(g(\rb_1,\rb_2) -1)\to (\nabla_1 n(\rb_1)) \frac{\rm const.}{|\rb_1-\rb_2|}(\nabla_2 n(\rb_2))$ is quite unclear.  First, how should the amorphous average density (a 'one point function') be inhomogeneous? Second, why are no long-ranged correlations in static two-point functions observed? It seems not easily possible to apply the concepts of DFT and equilibrium inhomogeneous density fields to glass, which is a metastable state.  Scattering experiments should detect the long-ranged density correlations accompanying this interpretation of glass as symmetry broken state \cite{Masters2001}, in the same way as the diffuse background diverges close to Bragg peaks in crystals \cite{Krivoglaz}. 

\subsection{Replica theory approach}

Another suggestion on how to derive the elasticity of glass from an interpretation of glass as a symmetry broken state was made by Szamel and Flenner \cite{Szamel2011}. They showed that in the mean-field replica approach, long-ranged displacement correlations, viz.~Eq.~\eqref{disprel}, and (shear) rigidity, viz.~finite elastic constants \eqref{def}, emerge from breaking of replica symmetry. In the replicated system, a glass shows up in non-trivial two-point  off-diagonal densities $n_{ij}(r)$, for different replicas $i\ne j$ of the system.   This overcomes the first problem in Sect.~4.1, and keeps the average density  $n(\rb)=n_0$ constant in glass. The long-ranged density correlations, expected following Wagner's Eq.~\eqref{bogol}, then appear in four-point replica off-diagonal densities. This eliminates the second problem in Sect.~4.1, and predicts finite correlations of the measurable densities. Of course, glass is an isotropic solid, so that the tensor of elastic constants simplifies to the expression containing the  two Lam\'{e} coefficients  $\lambda$ and $\mu$ only,
$C_{\alpha\beta\gamma\delta}=\lambda \delta_{\alpha\beta}\delta_{\gamma\delta}+\mu(\delta_{\alpha\gamma}\delta_{\beta\delta}+
\delta_{\alpha\delta}\delta_{\beta\gamma})$, where $\mu$ vanishes in a fluid.
Szamel and Flenner's final equation for the shear modulus $\mu$ closely resembles Eq.~\eqref{Cresult}, yet it contains replica off-diagonal densities  $n_{i\ne j}(r)$  and direct correlation functions $c_{i\ne jk\ne l}(\rb_1,\ldots,\rb_4)$ which can be evaluated with the replica Ornstein-Zernike equation \cite{Szamel2011}:
\begin{eqnarray}\label{mu}
\mu = \frac{k_B T}{2V} \int\!\!\! d\mathbf{r}_1 ... d\mathbf{r}_4 \;
y_{13}^2
\frac{\partial n_{10}(r_{12})}{\partial{x_1}}
\frac{\partial n_{10}(r_{34})}{\partial x_3}
\left[
c_{1010}(\rb_{12;}\rb_{34}) - c_{1020}(\rb_{12},\rb_{34})
\right].\nonumber
\end{eqnarray}
This approach thus achieves to combine the concept of symmetry breaking, spatially  correlated Goldstone modes, and elastic constants with the observed absence of these phenomena in the simple one- and two-point static structure functions of glass.   
The connection to the replica theory for the formation of a polymer gel \cite{Zippelius2004} appears interesting, as there the order parameter also is a second moment of a static density fluctuation which itself has vanishing average.  Comparing both approaches might provide additional information on the connection between Goldstone-modes and elasticity in amorphous solids.

\subsection{Mode coupling theory approach}

A practical approach to the moduli and elastic dispersion relations  in glass is inspired by the one for crystals with immobile defects: To determine the displacement field using the center of mass of the particle trajectory as lattice site \cite{Keim2004}; see the upper row of Fig.~3  for the crystal application (with mobile defects). The trajectory has to be averaged for a sufficient long time window but still short compared to the structural relaxation time $\tau_\alpha$ in order to test glassy not fluid-like behavior \cite{klix}. The theoretical analysis supporting this approach, derives a quasi-equilibrium description of the non-ergodic glass state, which is based on the assumption of a kinetic (not thermodynamic) glass transition. In this sense, the approach is inspired by mode coupling theory (MCT) \cite{goetze}, even though no explicit mode coupling approximations are involved. 

\begin{figure}[!ht]
	\includegraphics[width=.45\linewidth]{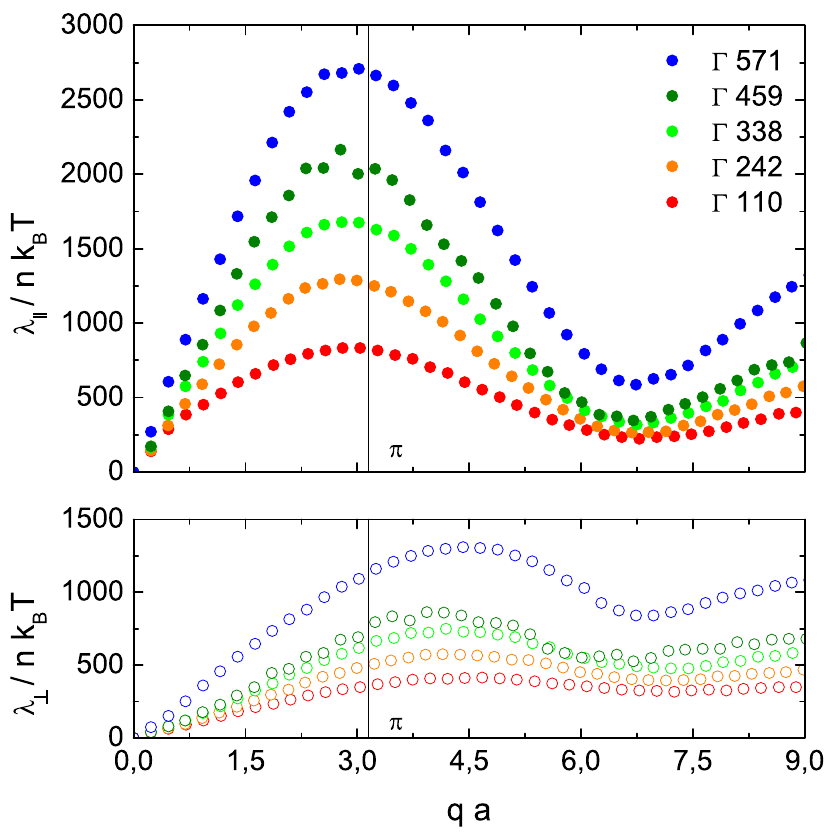}\hspace*{0.5cm}
	\includegraphics[width=0.45\textwidth]{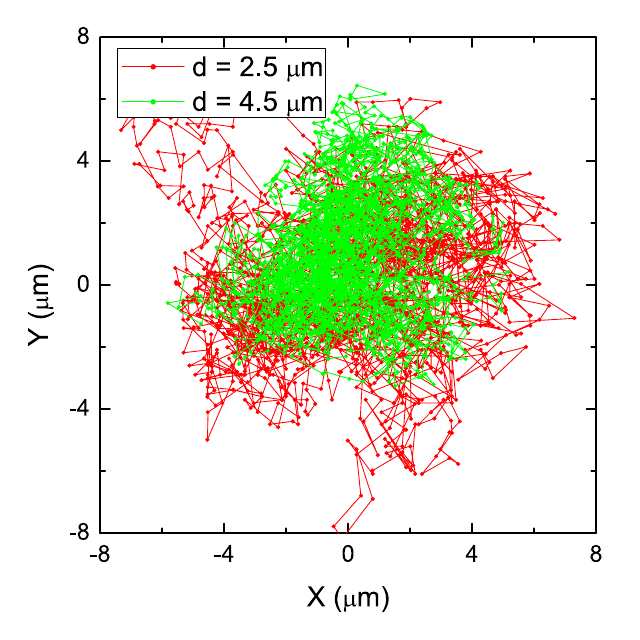}
\caption{The dispersion relation of 2D binary glass of soft repulsive colloids for different rescaled inverse temperatures $\Gamma$ as extracted from the displacements of particles from their time-averaged positions \cite{klix}; the averaging time $\Delta t\approx 10^4 s\ll \tau_\alpha$ is short enough so that starting from slightly below the glass transition $\Gamma_c\approx 200$ elastic behavior can be seen at intermediate times. (The average particle distance is $a\approx 21\mu m$.). Filled and empty symbols represent longitudinal and transverse waves, respectively. The right panel shows  typical particle trajectories of a big (green) and a small (red) particle at $\Gamma=242$, shifted to  center around the origin.}\label{fig01}
\end{figure}

The  defining relation which introduces the collective displacement field $\ub_\qb(t)$ is
\beq{e6m}
  \dot \ub_\qb(t)  = \vb_\qb(t)\; .
\eeq
Here the velocity field  $\vb_\qb(t)=(1/\sqrt{N}) \sum_{i=1}^N \; e^{ i \qb\cdot \rb_i(t)}\; \dot\rb_i(t)$ is well familiar from fluid hydrodynamics. (Note the convention to normalize fluctuations by $1/\sqrt{N}$, which is used in Sect.~4.3. The relation to the momentum density of Sect.~3.2 is $\hat{\bf j}=\sqrt{N} m \vb$.)
The central question is how to integrate Eq.~\eqref{e6m} over time. Let time averaging  be denoted by an overbar so that $\overline{r_i}$ is the average position of particle $i\in\left[1,N\right]$ during the time interval $\Delta t$. Then a possible ansatz for the particle's displacement is $\ub_i(t)=\rb_i(t)- \overline{r_i}$, and Eq.~\eqref{e6m} leads to
\begin{equation}\label{eq0}
\ub_\qb(t) =\frac{1}{\sqrt N}\,
\sum_{i=1}^N \;
e^{ i \qb\cdot \overline{\rb_i}}\;  \ub_i(t)\; ,
\end{equation}
to order  ${\cal O}(\qb\!\cdot\! \ub_i)$. Yet does this result  make sense? Do the displacements  remain bounded so that Eq.~\eqref{eq0} remains valid for a sufficient window in time?  Goldhirsch  and Goldenfeld solved Eq.~\eqref{e6m} differently and used $\ub_i(t)=\rb_i(t)- \rb_i(0)$, where time $t=0$ is some arbitrarily chosen initial time \cite{Goldhirsch2002}. They considered granular systems and were not concerned that the equilibrium-like equipartition theorem \eqref{disprel} does not follow from their approach; rather $\langle  \ub_\qb^*\;  \ub_\qb \rangle=\langle  \ub_\qb(t=0)^*\;  \ub_\qb(t=0) \rangle =0$ follows. Yet the discussion in Sect.~3.4 showed that Eq.~\eqref{eq0}   fails in non-ideal crystals because of defect diffusion; recall that it was used in the upper row of see Fig.~3 and gave results deteriorating with $\Delta t$, the time span over which the trajectory was averaged, and over which   ${\cal O}(\qb\!\cdot\! \ub_i(t))$ is required to remain small. 

Under the assumption of an ideal glass transition (as described by MCT), it  can be shown that the variance of displacements remains bounded using the Zwanzig-Mori equations of motion (EOM).  This justifies Eq.~\eqref{eq0} as leading order approximation to the solution of  Eq.~\eqref{e6m}. 
An approach considering glass as a non-ergodic state best starts in the ergodic state and considers simple and well defined correlation functions, which can become non-ergodic, viz.~take finite values at infinite time.  The following mean squared collective displacement function is a straightforward collective generalization of the single particle mean squared displacement:
\beq{e7}
{\bf C }(\qb,t) =   \langle \Delta \ub_\qb^*(t)\,\Delta \ub_\qb(t)  \rangle = 2 \int_0^t\!\!\! dt'\;(t\!-\!t')\; \langle \vb_\qb^*(t')\; \vb_\qb(0) \rangle \,,\eeq
where the displacement difference is $\Delta \ub_\qb(t)  = \int_0^tdt'\; \vb_\qb(t')$.
Its overdamped EOM appropriate for colloidal particles neglecting hydrodynamic interactions can be found from standard Zwanzig-Mori steps \cite{klix}:
\beq{e9}
{\bf C}(\qb,t)+ \frac{D_0 q^2}{k_BTn}\; \int_0^tdt'\; \tilde{\bf G}(\qb,t-t') \; {\bf C}(\qb,t') = 2D_0\, t\;  {\bf 1}\;\; ,
\eeq
with the short time diffusion coefficient $D_0$. Stress kernels generalize the (inverse) fluid isothermal compressibility $\kappa^T$ to finite wave vectors and frequencies:
$\tilde{\bf G}(\qb,t) = {\bf G}(\qb,t) + (1/\kappa_q^T) \hat{\qb}\hat{\qb}$. Here,  $\kappa^T_q=S_q/(k_BTn)$ is given by the equilibrium fluid structure factor.
The time-dependent stress kernels (with reduced dynamics according to the ZM-formalism \cite{Forster} indicated by $t_{\pQ}$)
 \beq{e8a}
{\bf G}(\qb,t)=(n/k_BT) \langle \sigb_{\qb}(t_{\pQ})^* \sigb_{\qb} \rangle\; ,\quad\mbox{with }\; 
\sigb_{\qb} = \frac{i}{q \sqrt{N}}\; \sum_{j=1}^N\;  \Fb_j
\;e^{i\rb_j \cdot {\bf q}}
\eeq
reduce to the stress auto-correlation functions in the limit of vanishing wave vector, with $\sigb_0$  a component of the potential part of the microscopic Irving-Kirwood stress tensor defined in Eq.~\eqref{kirkwood}:
\beq{e8b}
{\bf G}(t) = \lim_{q\to0} {\bf G}_\qb(t)  =  \frac{n}{k_BT}\; \langle  \sigb_0(t)^* \;\sigb_0  \rangle
\; .\eeq
The shear element  is measured as function of frequency $\omega$ in linear rheology, $G'(\omega)+iG''(\omega) = i \omega \int_0^\infty dt \; e^{-i\omega t} G(t)$, where $G'$ is the storage and $G''$ the loss modulus \cite{Siebenbuerger2012}.

Considering glass a non-ergodic state, the stress kernels like density (but not velocity) fluctuations  \cite{goetze} do not relax to equilibrium but take finite values at infinite time:
\beq{e10}
{\bf G}(\qb,t\to\infty) \to {\bf G}_\infty(\qb)\;,
\eeq
which shows up as finite zero-frequency elastic shear modulus in  rheology, $G'(\omega\to0)\to G_\infty$.  
Equation \eqref{e10} predicts using Eq.~(\ref{e9})
\beq{e11}
{\bf C}(\qb,t\to\infty) \to  {\bf C}_\infty(\qb) = 2 \frac{k_BT n}{q^2} \left({\bf G}_\infty(\qb)\right)^{-1}\; .
\eeq
Displacement differences stay below a finite limit for all times. Equation \eqref{eq0} is a valid approximation of displacement fluctuations in glass! The expansion $e^{ i \qb\cdot \rb_i(t)}=e^{ i \qb\cdot \bar{\rb}_i}+{\cal O}(\qb\cdot\ub_i(t))$ is justified, based on the collective variance as measure of the finite variance of individual particle displacements.  
This is an idealization as it requires that  $\tau_\alpha\gg \Delta t$, which restricts the approach to states (far) away from the transition (deep) in the glass.  Considering the local motion of the particles around the time-averaged positions, the equipartition theorem \eqref{disprel} follows and the dynamical matrix  ${\cal  D}(\mathbf{q})$ is given by the frozen-in stress fluctuations \cite{klix}
\begin{equation}\label{eq1}  \mathbf{\cal D}(\qb) = \frac{q^2}{n} \left( {\bf G}_\infty(\qb) + \frac{\hat{\qb}\hat{\qb}}{\kappa_q^T}  \right)  \, .
\end{equation}
Thus, all signatures of isotropic elasticity arise from treating glass as a metastable state with frozen-in stresses. 
The conservation law of momentum, which leads to the explicit factors $q^2$ in Eqs.~(\ref{e9}) and (\ref{eq1}), provides the route to explain the long-ranged nature of displacement fluctuations. 

Figure 4 shows dispersion relations in a two-dimensional colloidal system obtained by video microscopy \cite{klix}. Displacement fluctuations with time-averaged sites (see right panel), were correlated and the eigenvalues of the dynamic matrix in Eq.~\eqref{disprel} diagonalized. Previous microscopy approaches had concentrated on the covariance matrix to obtain the density of states \cite{Ghosh2010}. The dispersion relations display the signatures of a solid, viz.~transverse elastic modes,  in colloidal glass and in viscoelastic fluids at intermediate times; the latter aspect follows Maxwell's insight that fluids are solid-like at high frequencies \cite{Landau}. Considering that stresses are notoriously difficult to measure in dense amorphous systems, the smoothness of the dispersion relations obtained from the equipartition theorem is noteworthy. Computer simulations also explicitly tested relation \eqref{eq1}, and determined the Lam\'{e}  coefficient $\mu$ and the frozen-in shear stress $G_\infty$ independently. Agreement as predicted by Eq.~\eqref{eq1} was observed within the statistical error bars \cite{klix}. 

\section{Conclusions and outlook}

 This short review compared different theories for the displacement correlations, elastic dispersion relations, and elastic moduli, which can be studied by techniques like video or confocal microscopy in colloidal solids. The effects of disorder either as defects in periodic solids or in amorphous solids was highlighted. A number of theoretical relations, e.g.~between fluctuation and DFT results, remain to be understood, and experiments in three-dimensional and two-dimensional solids would be useful. Extensions to optical phonons in periodic solids, and phasonic modes in quasi-crystalline ones  \cite{Lifshitz2011} are promising. In glasses, the role of defect motion remains open, and extensions to capture the
Boson peak and the density of states \cite{Goetze2000} should be included in the theoretical approach. 
\vfill 

\appendix \section{}

The free energy density $f=F/V$  is derived starting from the free energy relation in Eq.~\eqref{firstlawF}  in order to  determine its dependence on the  thermodynamic fields; one finds  $f=f(u_{\alpha\beta},n)$, respectively $f=f(u_{\alpha\beta},c)$, with $u_{\alpha\beta}$ the strain, $n=N/V$ the total density, and $c$ the defect density  \cite{Martin1972,Chaikin95}. Constant temperature ($dT=0$) is considered.

 Because of the homogeneity of $F$ in the thermodynamic limit, one obtains
$$f=F/V= \mu \; n + h_{\alpha\beta}\; u_{\alpha\beta} - p(\mu,h_{\alpha\beta})\: .$$
Here the Gibbs-Duhem relation was already indicated, which states the dependence of the pressure $p$ on chemical potential $\mu$ and stress tensor $\bf h$:
$$ d p = n\; d\mu +  u_{\alpha\beta}\; d h_{\alpha\beta}\;.$$
It follows from $F\!-\!\mu N\!-\!{\rm Tr}[{\bf h\!\cdot\! U}]= - p V$ being extensive  with $V$ as single extensive variable. 
Combining the total differential of the first equation with the second gives
$$d f = \mu \; dn  +  h_{\alpha\beta}\; d u_{\alpha\beta}\;,$$
the dependence of the free energy density on strain and total density.  Transforming to defect density $c$ via $dn= - n du_{\alpha\alpha}-dc$, gives the free energy density as function of  strain and defect density:
$$d f = - \mu \; dc  + \sigma_{\alpha\beta}\; d u_{\alpha\beta}\;,\quad\mbox{with }\;
 \sigma_{\alpha\beta}=  h_{\alpha\beta} - n \mu \; \delta_{\alpha\beta}\;.$$
This gives for the elastic constants $C_{\alpha\beta\gamma\delta}^{c}=\left.\partial \sigma_{\alpha\beta}/\partial u_{\gamma\delta}\right)_{c}$.

Fleming and Cohen rewrite $f$ in terms of the free energy density $\bar f=F/N$ per particle \cite{Fleming1976}  using $df=d\left( n \bar{f}\, \right) =n \,d \bar{f} +   \bar{f}\, dn$. This leads to
$$n\, d \bar{f} = - \frac pn \; dc  + t_{\alpha\beta}\; d u_{\alpha\beta}\;,\quad\mbox{with }\;
 t_{\alpha\beta}=  h_{\alpha\beta} - \left(n \mu - f \right)\; \delta_{\alpha\beta}\;.$$
The stress tensor $\bf t$ gives the reversible part of the momentum conservation equation \cite{Walz2010,Fleming1976}, which shows that  it can be obtained from averaging the microscopic Irving-Kirkwood stress tensor in Eq.~\eqref{kirkwood}. For fixed $N$, the last equation recovers Eq.~\eqref{eq5}. In unstrained equilibrium where $h_{\alpha\beta}=0$, this confirms Eq.~\eqref{pressure} and gives the connection $ C_{\alpha\beta\gamma\delta}^{(t)}=C_{\alpha\beta\gamma\delta}^{c}-n\mu\delta_{\alpha\beta}\delta_{\gamma\delta}$.
\bigskip

\acknowledgments 
{\it Acknowledgements}: 
Parts of the reviewed results are own work, which was obtained with C. Walz, G. Szamel, C. Klix, F. Ebert, F. Weysser, G. Maret, and P. Keim. I cordially thank them for the very fruitful cooperations.  Also, I would like to acknowledge U. Gasser,  F. Miserez and Th. Voigtmann for helpful discussions, C. Walz and G. Szamel for a critical reading of the manuscript, and U. Gasser  for preparation of Figures 3. The work was (partly) funded by the German Excellence Initiative. 

\bibliography{lit}
\bibliographystyle{varenna}
\end{document}